\documentclass[aps,pra,twocolumn,groupedaddress,amssymb,amsmath]{revtex4-1} 

\usepackage[percent]{overpic} 
\usepackage{xcolor} 
\usepackage{upgreek}
\usepackage{graphicx}
\usepackage{amsmath}
\usepackage[hidelinks]{hyperref}
\usepackage[normalem]{ulem}

\newcommand{\textold}[1]{\textcolor{red!90!black}{\sout{#1}}}
\newcommand{\textnew}[1]{\textcolor{green!50!black}{#1}}

\renewcommand{\textold}[1]{}
\renewcommand{\textnew}[1]{#1}

\begin{document}

\title{Dilute dipolar quantum droplets beyond the extended Gross-Pitaevskii equation}

\author{Fabian B\"{o}ttcher}
\author{Matthias Wenzel}
\author{Jan-Niklas Schmidt}
\author{Mingyang Guo}
\author{Tim Langen}
\author{Igor Ferrier-Barbut}
\altaffiliation[Present address: ]{Laboratoire Charles Fabry, Institut d’Optique Graduate School, CNRS, Université Paris-Saclay, 91127 Palaiseau Cedex, France}
\author{Tilman Pfau}
\email{t.pfau@physik.uni-stuttgart.de}
\affiliation{5{.} Physikalisches Institut and Center for Integrated Quantum Science and Technology, Universit{\"a}t Stuttgart, Pfaffenwaldring 57, 70569 Stuttgart, Germany}

\author{Ra\'{u}l Bomb\'{i}n}    
\author{Joan S\'{a}nchez-Baena}
\author{Jordi Boronat} 
\author{Ferran Mazzanti} 
\email{ferran.mazzanti@upc.edu}
\affiliation{Departament de F\'{i}sica, Universitat Polit\`{e}cnica de Catalunya, Campus Nord B4-B5, E-08034, Barcelona, Spain}
\date{\today}

\begin{abstract}
Dipolar quantum droplets are exotic quantum objects that are self-bound due to the subtle balance of attraction, repulsion and quantum correlations. Here we present a systematic study of the critical atom number of these self-bound droplets, comparing the experimental results with extended mean-field Gross-Pitaevskii equation (eGPE) and quantum Monte-Carlo simulations of the dilute system. The respective theoretical predictions differ, questioning the validity of the current theoretical state-of-the-art description of quantum droplets within the eGPE framework \textnew{and indicating that correlations in the system are significant}. Furthermore, we show that our system can serve as a sensitive testing ground for many-body theories in the near future.
\end{abstract}
    
\maketitle

\section{Introduction}
    
For systems with competing interactions, quantum fluctuations can stabilize an otherwise collapsing system \cite{Petrov2015}. The balance of attraction and repulsion in these systems means that they share properties with liquids, despite being orders of magnitude more dilute. These quantum droplets were experimentally discovered by driving a dipolar Bose-Einstein condensate (BEC) into the strongly dipolar regime. However, instead of collapsing, the system formed stable droplets \cite{Kadau2016}. Since their discovery, many properties of dipolar quantum droplets have been observed and compared to the predictions of their current state-of-the-art theoretical description, the extended Gross-Pitaevskii equation (eGPE). These predictions include the stabilization of the droplets typically explained by the Lee-Huang-Yang (LHY) correction of the mean-field energy \cite{Ferrier-Barbut2016, Chomaz2016a}, their self-bound nature \cite{Schmitt2016a}, collective modes \cite{Chomaz2016a, Ferrier-Barbut2018a}, the emergence of striped states in confined geometries \cite{Wenzel2017}, as well as the existence of arrays of phase coherent droplets with transient supersolid properties \cite{Boettcher2019, Tanzi2018, Chomaz2019}. In addition to dipolar systems, quantum droplets have also been observed in Bose-Bose mixtures \cite{Cabrera2017, Semeghini2018, Cheiney2018, Ferioli2018}. 

Here, we systematically study the critical atom number that is necessary to form the liquid-like droplet state, extending previous results with $^{164}$Dy \cite{Schmitt2016a} to an order of magnitude higher atom number. Comparing the measured critical atom number to results obtained from numerical simulations of the eGPE seems to indicate a systematic shift to lower values. Experimental discrepancies compared to the eGPE predictions have also been observed in other related systems \cite{Petter2018, Cabrera2017, Chomaz2019}. 

Motivated by this observation we present a theoretical approach that goes beyond the framework of the eGPE. We solve the dilute many-body system using quantum Monte-Carlo calculations (QMC), in particular the Path Integral Ground State (PIGS) method~\cite{Sarsa_2000, Rota_2006}. With this we can extract the critical atom number of self-bound droplets, in good agreement with the experimental measurements. As a mean field theory, the eGPE framework is limited to the usage of the local density approximation, as well as to the perturbative regime at small gas parameters \cite{Gautam2018}. In contrast our PIGS calculations intrinsically include effects due to the finite system size and of the finite interaction range, as well as particle correlations and quantum fluctuations. With our method we directly have access to the correlations in the system, which we use to extract the spatial pair correlation function, as well as the condensate depletion, which is increased compared to the prediction of Bogoliubov theory. These results suggest that in the density regime relevant for quantum droplets, the state-of-the-art eGPE framework is not able to reproduce all observable properties of these quantum droplets.

\section{Experiment}

For the experiments we use $^{\text{162}}$Dy with a magnetic dipole moment $\mu = 9.93\,\mu_{\text{B}}$, where $\mu_{\text{B}}$ is the Bohr magneton. To quantify the relative strength of the dipole-dipole interaction compared to the contact interaction, we define the relative dipolar strength $\varepsilon_{\text{dd}} = a_{\text{dd}}/a_{\text{s}}$, in terms of the scattering length $a_{\text{s}}$ and the dipolar length $a_{\text{dd}} = \mu_{\text{0}}\mu^2 m/(12 \pi \hbar^2)$. Here $\hbar$ is the reduced Planck constant, $\mu_{\text{0}}$ is the vacuum permeability and $m$ is the atomic mass. The dipolar length is different for the two studied isotopes and has a value of $129\,a_{\text{0}}$ for $^{162}$Dy, and $131\,a_{\text{0}}$ for $^{164}$Dy, with the Bohr radius $a_{\text{0}}$. Like all lanthanide atoms, $^{\text{162}}$Dy exhibits a rich spectrum of Feshbach resonances that can be used to control the strength of the short-range contact interaction \cite{Chin2010, Baumann2014, Frisch2014, Maier2015, Maier2015a}. Here, we use a specific double resonance at around 5.1\,G \cite{Baumann2014, SupMat} that, together with the high background scattering length $a_{\text{bg}} = 140(20)\,a_{\text{0}}$ of $^{\text{162}}$Dy \cite{Tang2018, Tang2015a, Tang2016}, allows us to tune the scattering length $a_{\text{s}}$ from a contact-dominated sample (away from the resonances), to a dipolar-dominated sample closer to the zero-crossing of the scattering length. Since the background scattering length of $^{\text{162}}$Dy has so far not been determined to high precision, all calculated scattering lengths exhibit a systematic uncertainty of 15\%. 

While a BEC is a gaseous state, which means that it freely expands in the absence of an external trap, the droplet state is self-bound due to its intrinsic interactions \cite{Baillie2016,Wachtler2016a, Schmitt2016a, Cikojevic2018, SupMat}. This self-bound character has been experimentally observed for $^{164}$Dy \cite{Schmitt2016a}, as well as for the related quantum droplets in Bose-Bose mixtures \cite{Cabrera2017, Semeghini2018}. In order to probe this feature experimentally, we initially prepare a quasi-pure BEC with $4.5(3)\times10^4$ $^{162}$Dy atoms and then closely follow the procedure presented in \cite{Schmitt2016a, SupMat} to create a single self-bound droplet. After a variable evolution time we intentionally evaporate the droplet and subsequently image it after an expansion of either 8\,ms or 30\,ms, depending on the atom number. With this approach we observe atom number decay curves that settle at a constant atom number -- the critical atom number of a self-bound droplet \cite{Schmitt2016a, SupMat}.
 
\begin{figure}[t]
\begin{overpic}[width=0.48\textwidth]{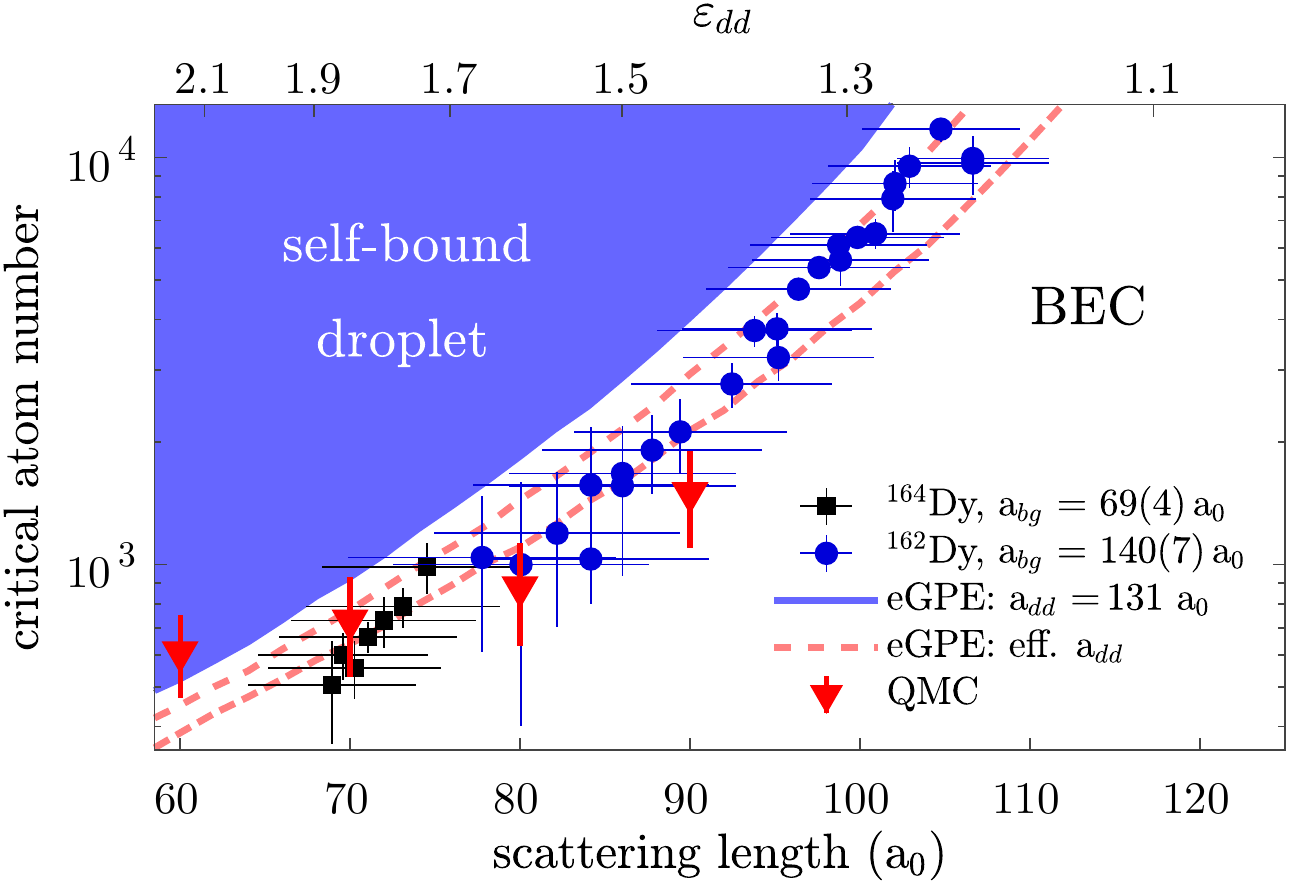}
\end{overpic}
\vspace{-3mm}
\caption{\label{fig:critNum}
Critical atom number of a self-bound dipolar quantum droplet for $^{162}$Dy (blue points) and $^{164}$Dy \cite{Schmitt2016a} (black squares). We extract the critical atom number by analyzing atom number decay curves \cite{SupMat}. The theoretical boundary of the phases is obtained from numerical eGPE simulations. The red dashed and dash-dotted lines show the corresponding boundary as expected from an increased effective dipolar length due to a finite collision energy of 30-50 and 100\,nK, respectively \cite{Odziejewski2016, *Odziejewski2017Erratum, JachymskiPrivateComm}. The red triangles show the results obtained by quantum Monte-Carlo (QMC) simulations, with the error bars chosen to cover the uncertainties of both the statistical error and the non-universality. See main text for more information.}
\end{figure}

We want to compare our results to numerical eGPE and Quantum Monte-Carlo simulations, as well as previous measurements with $^{164}$Dy published in \cite{Schmitt2016a}. The two bosonic isotopes of dysprosium, $^{162}$Dy and $^{164}$Dy, only differ in their mass, which manifests in a shift of $m_{164}/m_{162} = 1.2\%$ of the dipolar length, and thus on the scattering length axis. This effect is smaller than our uncertainty in this quantity, and we can thus safely neglect it. For the comparison with the $^{164}$Dy measurements we use the Feshbach resonances discussed in \cite{Kadau2016, Schmitt2016a} together with the latest measurement of the background scattering length in the droplet state $a_{\text{bg,164}} = 69(4)\,a_{\text{0}}$ \cite{Ferrier-Barbut2018a}. With this we observe systematically lower critical atom numbers compared to the eGPE simulations, that could be accounted for by the uncertainty of the scattering length calibration. For our new results, we first use the literature value of the background scattering length $a_{\text{bg,162}} = 140(20)\,a_0$ \cite{Tang2018, Tang2015a, Tang2016}, together with our measured parameters of the Feshbach resonances \cite{SupMat}, leading to a similar systematic shift to lower atom numbers as in the $^{164}$Dy data. Again, this shift could be explained by the even larger uncertainty in the $^{162}$Dy scattering length.

Next, we use the sensitive scaling of the critical atom number with respect to the scattering length in order to extract the ratio of the two background scattering lengths $a_{\text{bg,162}}/a_{\text{bg,164}}$ \cite{SupMat}, free of the systematic uncertainties of their respective measurements. To put this scaling into context, we note that the current uncertainty of the background scattering length of $^{162}$Dy is about 20\,$a_{\text{0}}$. The critical atom number changes by an order of magnitude over a comparable range of the scattering length, showing that our measurement constitutes a very sensitive probe of the scattering length. Here, we use the critical atom numbers derived from our eGPE simulations to extract the ratio $a_{\text{bg,162}}/a_{\text{bg,164}}$. Naturally, the same procedure can be done using any quantum many-body theory able to predict critical atom numbers for a self-bound droplet.

Starting from the latest measurement of the background scattering length of $^{164}$Dy \cite{Ferrier-Barbut2018a} we first shift the eGPE critical number curve \footnote{Note, that this procedure assumes that the critical atom number only depends on the scattering length and not on other parameters, e.g. the actual density distribution of the droplet, as well as the scaling extracted with the eGPE framework.} in order to minimize the difference to the experimental data for $^{164}$Dy \cite{SupMat}. From this point on we optimize the background scattering length $a_{\text{bg,162}}$ to minimize the residual of our new measurements with respect to this shifted theory curve \cite{SupMat}. Taking into account the residual systematic uncertainty of the background scattering length of $^{164}$Dy \cite{Ferrier-Barbut2018a}, we end up with $a_{\text{bg,162}} = 140(7)\,a_0$, in agreement with the literature value \cite{Tang2018, Tang2015a, Tang2016}, but with a significantly reduced uncertainty. More importantly, since it is independent of this residual systematic uncertainty, we can extract the ratio of the two background scattering lengths. Comparing the two isotopes we find a ratio of $a_{\text{bg,162}}/a_{\text{bg,164}} = 2.03(6)$ of their respective background scattering lengths. 

This way we calibrate the scattering length of $^{162}$Dy  and show a summary of all measured critical atom numbers in Fig.~\ref{fig:critNum}. The experimental atom number uncertainties are chosen to cover both the determination of $N_{\text{crit}}$ \cite{SupMat}, with an additional 10\% uncertainty of the imaging. The uncertainty of the scattering lengths are based on our experimental magnetic field stability of $\sim$2\,mG, the knowledge of the used Feshbach resonances, as well as the uncertainty of the respective background scattering length. 

Experimentally we seem to find systematically lower critical atom numbers compared to eGPE predictions. The straightforward experimental source would be a shift of approximately 6\,$a_0$ in the scattering length axis. This is at the edge of the error bars for $^{164}$Dy, and within the error bars of the independent calibration for $^{162}$Dy. Note that our calibration of the $^{162}$Dy scattering length is consistent with our recent experimental observation of supersolid properties in an array of $^{162}$Dy quantum droplets \cite{Boettcher2019}, which agrees with theoretical calculations in a narrow window of the scattering length. In order to make the presented measurements a sensitive benchmark for quantum many-body theories, an independent and precise measurement of the scattering length $a_{\text{s}}$ would be required for the two isotopes. Another possible experimental explanation would be a systematic uncertainty in the atom number determination, which we rule out by performing measurements with independent imaging techniques, all resulting in similar values within the quoted error bars.

\section{Theory}

We now focus on possible theoretical explanations of systematically lower critical atom numbers of the self-bound state. Within the eGPE framework, it was shown in \cite{Odziejewski2016, *Odziejewski2017Erratum} that it is possible to account for finite temperature effects in the two-body scattering problem by effectively enhancing the dipolar length $a_{\text{dd}}$. In Fig.~\ref{fig:critNum} we also show two theoretical curves for an enhanced dipolar length corresponding to a collision energy of 30-50\,nK \cite{JachymskiPrivateComm} (dashed red line), and 100\,nK (dash-dotted red line). The resulting shift for the critical atom number is in good agreement with the experimental results, suggesting that such an effective enhancement of the dipolar length might play an important role in dipolar scattering at finite collision energies. In the future, we plan to resort to spectroscopic measurements on embedded impurities \cite{Wenzel2018} in order to directly measure the temperature of these quantum droplets and therefore determine the actual strengths of this proposed correction.

\begin{figure}[t!]
\begin{center}
\begin{overpic}[width=0.48\textwidth]{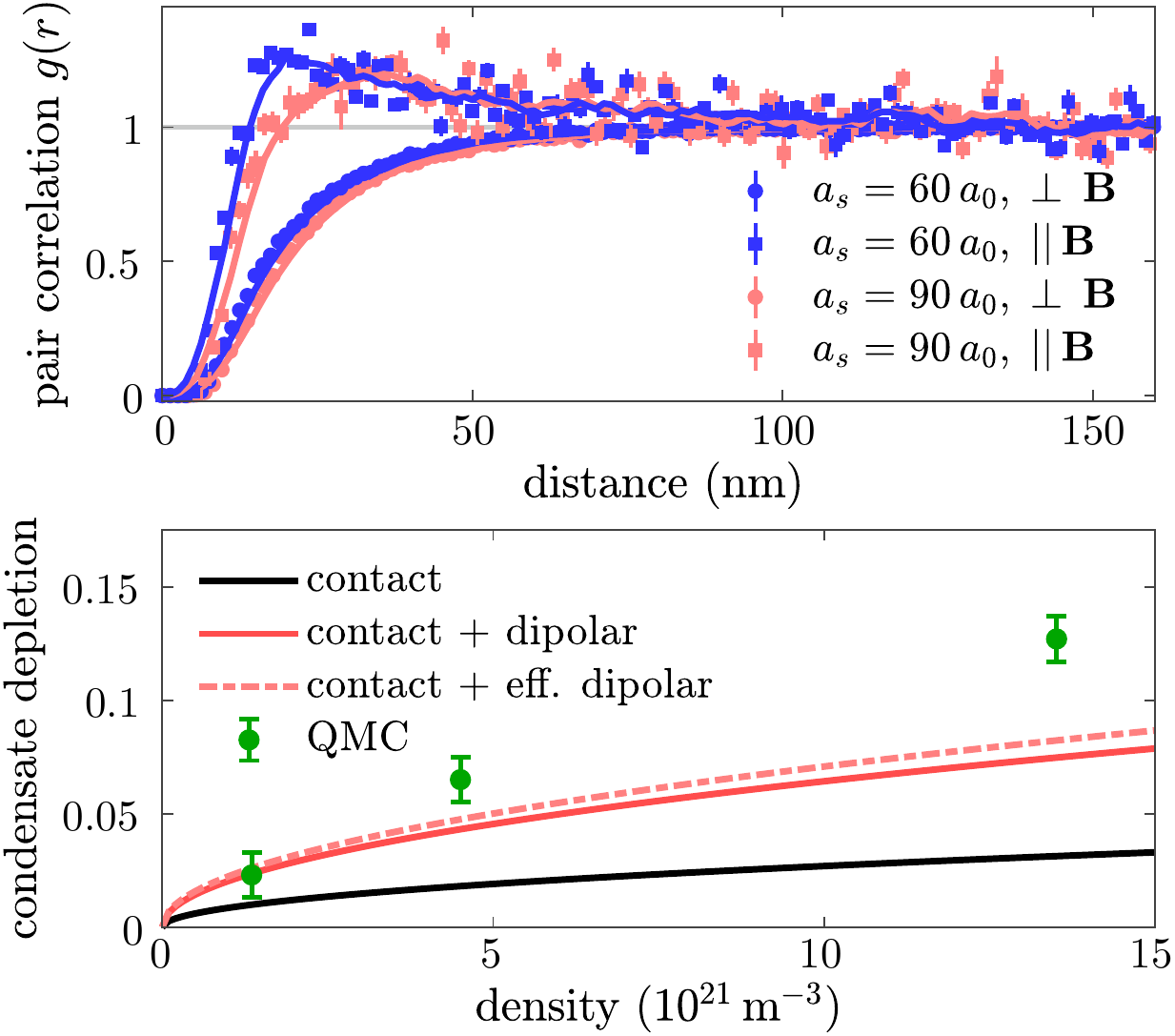}
\put(93.5, 84){a)} \put(93.5, 39){b)}
\end{overpic}
\caption{(a) Pair distribution function $g(\textbf{r})$ for the bulk system at a density of $n=5.88\times10^{21}\,$m$^{-3}$, corresponding to the central density of a saturated quantum droplet at $a_s=60\,a_0$. The red (blue) symbols correspond to a scattering length of $a_s=60\,a_0$ ($a_s=90\,a_0$), while the squares (points) indicates the direction along (perpendicular to) the polarization direction. The solid lines act as a guide to the eye.
(b) Condensate depletion as predicted by the PIGS calculations and the Bogoliubov theory -- without and with dipolar interaction -- for a scattering length of $a_s=60\,a_0$.}
\label{fig:correlations}
\end{center}
\end{figure}

Next, we present an approach that goes beyond the current state-of-the-art eGPE framework. For this we solve the dilute many-body system at zero temperature using quantum Monte-Carlo simulations. For weakly interacting systems, the results obtained in QMC are in good agreement with mean-field predictions \cite{Macia2011, Saito2016, Giorgini1999, Mazzanti2003}. However, for strongly correlated systems, e.g. liquid helium \cite{Ceperley1995, Boronat2002}, mean-field fails, while QMC still provides reliable predictions. To see whether correlations influence the properties of quantum droplets, we now focus on QMC simulations \cite{Saito2016, Macia2016, Bombin2017}. In particular, we use the PIGS method \cite{SupMat} to determine the ground-state properties of ensembles of $^{162}$Dy atoms at zero temperature. Although computationally extensive, this method intrinsically includes finite range effects present in a more realistic description of the atom-atom interaction, the finite system size and a correct description of correlations and quantum fluctuations. Compared to this, the eGPE approach relies on the local density approximation for the quantum fluctuations described by the LHY term \cite{Lee1957, Lima2011}. Our approach is, however, limited to the usage of an effective Hamiltonian without bound states to describe the interaction between particles. This effective Hamiltonian \cite{SupMat} includes the dipolar interaction and a central two-body interaction, with a repulsive core and a realistic $C_6$ coefficient \cite{Li2017, SupMat}. In order to study whether the system is universal, we use different model potentials \cite{SupMat}, with the respective parameters fixed to adjust the zero-energy $s$-wave scattering length to the range of the experimentally measured values. This is accomplished by solving the Lippmann-Schwinger equation associated to the $T$-matrix of the full interaction.

First, we study the influence of correlations compared to the Bogoliubov prediction used to derive the LHY term \cite{Lee1957, Lima2011} in the eGPE framework. Therefore, we simulate the equivalent homogeneous bulk system with a density of $n=5.88\times10^{21}\,$m$^{-3}$ for two different scattering lengths, $a_s=60\,a_0$ and $a_s=90\,a_0$, and extract the pair correlation function $g(\textbf{r})$. The density is chosen to correspond to the equilibrium density of a saturated droplet at $a_s=60\,a_0$. Due to the anisotropy of the dipolar interaction, the correlation function $g(\textbf{r})$ shown in Fig.~\ref{fig:correlations}(a) depends on the direction with respect to the polarization axis. Perpendicular to the polarization axis, the pair correlation function is a monotonic function of the distance that resembles the one of a weakly interacting system. On the other hand, along the polarization direction it shows signatures of local ordering, as highlighted by a broad peak at short distances. In both directions, the length scale of the repulsion at short distances is caused by the repulsive core of the used two-body model potential.

Another property that directly measures the strength of correlations, is the quantum depletion $1 - n_c/n$, with the condensate fraction $n_c/n$. The condensate fraction equals one for an ideal Bose gas at zero temperature, and decreases when correlations are enhanced. In strongly interacting liquid helium the condensate fraction is typically below 10$\%$ \cite{Glyde2011}. For typical weakly interacting ultra-cold atom experiments it is around $99\%$, while by largely increasing the scattering length, condensate fractions as low as about 90\% \cite{Lopes2017} have been realized. In Fig.~\ref{fig:correlations}(b) we show the comparison between the PIGS prediction and the derived quantum depletion $1- n_c/n$ within the Bogoliubov theory for the weakly interacting Bose gas with no dipolar interaction, as well as with dipolar interaction \cite{Lee1957, Lima2011}, for the corresponding parameters of our quantum droplets. As it can be seen, the dipolar interaction leads to stronger correlations, as well as to a larger overall depletion of the condensate in the range of densities relevant for saturated quantum droplets. Compared to this, the PIGS results show that the effect of correlations is even stronger. 

Now, we turn to the study of the full, finite system. We analyze realizations of the system with different number of particles for a given $s$-wave scattering length and compute the ground state energy in each case. Like for the eGPE simulations, we identify a self-bound droplet as a system with negative energy in the absence of an external trapping confinement. The self-bound droplets predicted by our PIGS calculations differ from those obtained in the eGPE approximation in the overall density profiles \cite{SupMat}. This difference can be attributed to the presence of correlations, which can be quantified by looking at the two-body properties shown in Fig.~\ref{fig:correlations}. By looking at the density distributions of self-bound droplets, we therefore plan to distinguish the two proposed theories experimentally in the future. As in the experiments and the eGPE simulations, we find that there is a critical atom number below which the system ceases to be self-bound. Close to the critical atom number our PIGS calculations result in a lower peak density than predicted within the eGPE framework \cite{SupMat}. Note, that there is a non-negligible dependence of the critical atom number on the exact model of the two-body potential we employ, which indicates that the problem is non-universal in terms of the scattering length. The resulting critical atom numbers for several values of the scattering length are shown with red triangles in Fig.~\ref{fig:critNum}, with the error bars chosen to take into account the effect of the non-universality according to the analyzed model potentials, as well as the statistical errors of the simulations. The obtained critical atom numbers are always below the eGPE predictions and in good agreement with the experimental measurements. The improvement of the PIGS predictions with respect to the eGPE results points to the relevance of finite-range effects which enhance quantum correlations, similar as in dilute Bose mixtures \cite{Staudinger2018}.

\section{Conclusion}

In conclusion we have systematically studied the critical atom number for a self-bound dipolar quantum droplet experimentally, and have used these measurements to establish a comparison between current state-of-the-art eGPE description and quantum Monte-Carlo simulations. Compared to eGPE \textold{results} results, we observe indications of a systematic shift of the experimentally measured critical atom numbers to lower values.
\textnew{Those values are nevertheless well reproduced by a zero--temperature quantum Monte Carlo simulation based on the PIGS algorithm. In contrast to the eGPE, our PIGS calculations include finite range effects in the interaction as well as finite size effects, together with correlations and quantum fluctuations. This is used to extract the spatial pair correlation function and the condensate depletion, showing that in the relevant density regime of quantum droplets, correlations are enhanced and need to be included in a realistic description of the problem. In this way, our PIGS results indicate that correlations are beyond what a zero-temperature modified mean field theory can capture.  Alternatively, the inclusion of finite temperature effects in the eGPE framework (through an effective re-normalization of the dipolar interaction strength associated to finite collision energies in the two-body scattering problem), can also reproduce the critical number data, although the prediction is strongly dependent on the temperature used in the calculation.}
\textold{We show that one possibility to capture this behavior within the eGPE framework is to introduce an effective re-normalization of the dipolar interaction strength associated to finite collision energies in the two-body scattering problem. Moving beyond the eGPE framework, we have presented quantum Monte-Carlo simulations of the dilute many-body
system at zero temperature. In contrast to the eGPE simulations, our PIGS calculations intrinsically include the finite system size, effects of the finite range of the interactions, as well as correlations and quantum fluctuations. This is used to extract the spatial pair correlation function and the condensate depletion, showing that in the relevant density regime of quantum droplets, correlations are enhanced and need to be included.}
All in all, our results call into question the validity of the eGPE framework to fully describe the quantum droplets. \textold{In the future, we expect our measurements -- complemented by an independent and precise determination of the scattering length, as well as the temperature -- to serve as a sensitive testing ground for many-body theories.}

\begin{acknowledgments}
This work is supported by the German Research Foundation (DFG) within FOR2247 under Pf381/16-1, Pf381/20-1, and FUGG INST41/1056-1. T.L. acknowledges support from the EU within Horizon2020 Marie Sk\l odowska Curie IF (Grant~No.~746525~coolDips).
I.F.B. acknowledges support from the EU within Horizon2020 Marie Sk{\l}odowska Curie IF (Grant~No~703419~DipInQuantum). 
Partial financial support from the MINECO (Spain) grant No. FIS2017-84114-C2-1-P is also acknowledged. J. Sánchez Baena aknowledges the FPU fellowship with reference FPU15/01805 from MECD (Spain) R. Bombín aknowledges the FPI fellowship BES2015-074088 from MINECO (Spain).
\end{acknowledgments}


\appendix
    
\section{Experiment}

Our experimental setup creates Bose-Einstein condensates of either $^{164}$Dy or $^{162}$Dy in a crossed optical dipole trap (cODT) (along $\hat{x}$ and $\hat{y}$ axes, $\lambda_{\text{cODT}}~=~1064\,$nm). For the experiments with $^{162}$Dy we typically create a quasi-pure BEC consisting of $4.5(3)~\times~10^4$ atoms at a temperature below $T~\approx~16(3)$\,nK in a slightly oblate trap. For $^{162}$Dy, this is done by forced evaporation in the cODT away from the Feshbach resonances with $B_{\text{BEC}}~=~5.875\,\text{G}$ corresponding to  $a_{\text{s}}~\approx~a_{\text{bg}}$. After this we typically change the trap and/or the magnetic field for the actual experiments, as described in further detail for the performed experiments below.

Along the $\hat z$ axis we have a microscope objective allowing for in-situ imaging with $1\,\upmu$m resolution. We can use this microscope for far-detuned phase-contrast imaging as well as resonant absorption imaging. Both techniques, as well as an independent time-of-flight imaging along the $\hat{y}$-direction, result in similar atom numbers. The microscope can also be used to focus an additional optical dipole trap ($\lambda~=~532\,$nm) that has a calculated beam waist of $\sim~22\,\upmu$m, to change the trap aspect ratio from the oblate cODT to a spherical or even prolate trap.

\subsubsection*{Self-bound droplet measurements}

To measure the critical atom number for self-bound droplets shown in Fig.~\ref{fig:critNum} of the main text, as well as the measurement of the expansion velocity in Fig.~\ref{SupMatFig:ExpVelo}, we apply a magnetic field gradient along the $\hat{\textbf{z}}$ direction after the preparation of the BEC. This applied gradient exactly compensates gravitational forces and leads to a shift of the magnetic field by $-428\,$mG, that we compensate by ramping up the amplitude of the constant magnetic offset field at the same time. After this we reshape the trap within 20$\,$ms into a spherical trap with a mean trap frequency of $\bar{\omega}~=~96(4)$\,Hz. To do this we change the trap aspect ratio from $\lambda =  \omega_{\text{z}}/\omega_{\text{r}} = 96(2)\,\text{Hz}/25(2)\,\text{Hz} = 3.8$ to $\lambda = 98(2)\,\text{Hz}/94(4)\,\text{Hz} = 1.05$, by applying an additional optical dipole trap along the $\hat{z}$-direction. Here $ \omega_{\text{z}}$ ($\omega_{\text{r}}$) is the trapping frequency along (perpendicular) to the magnetic field direction. For such a spherical trap geometry the regular BEC at large scattering lengths and droplet phase at low scattering lengths are connected by a continuous crossover \cite{Ferrier-Barbut2016, Bisset2016, Wachtler2016a}, as can also be seen in the phase diagram shown in Fig.~\ref{SupMatfig:2}. To reach the droplet state we ramp the magnetic field amplitude within 20\,ms closer to the zero-crossing Feshbach resonance to reduce the scattering length. Subsequently, we let the atoms evolve for 10\,ms to allow the droplets to form. Then we slowly switch off the trap by ramping down the intensities of the laser beams to $\sim$5\% of their initial values, keeping a nearly constant trap aspect ratio. We then suddenly turn off the trap and levitate the cloud for various times before subsequently imaging the density distribution. 

For the measurements of the expansion velocity shown in Fig.~\ref{SupMatFig:ExpVelo} we follow the expansion up to $t_{\text{tof}} =$ 20\,ms and subsequently image the atoms using far-detuned phase contrast imaging. 

For the measurements of the critical atom number, as shown in Fig.~\ref{fig:critNum} of the main text, we levitate the atomic cloud for a variable evolution time and then intentionally evaporate the droplets \cite{Schmitt2016a} by ramping up the magnetic field within $100\,\upmu$s to $B \approx 6.0\,$G. At this field the ground state of the system is an expanding BEC and therefore the droplet evaporates back to the gaseous phase and expands freely. After an expansion of either 8\,ms or 30\,ms, depending on the atom number, we image the atomic cloud with resonant absorption imaging. With this we can observe atom number decay curves, that settle to a constant value -- the critical atom number of the self-bound droplet. The shown scattering length range in Fig.~\ref{fig:critNum} of the main text for $^{162}$Dy corresponds to magnetic fields in the range from $B$~=~5.293\,G to 5.249\,G.

\section{Feshbach resonances and three-body loss coefficient}  

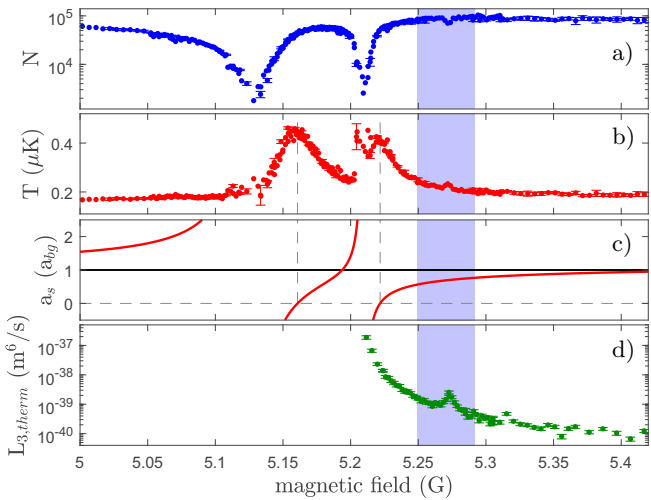
\begin{figure}[t]
\begin{overpic}[width=0.48\textwidth]{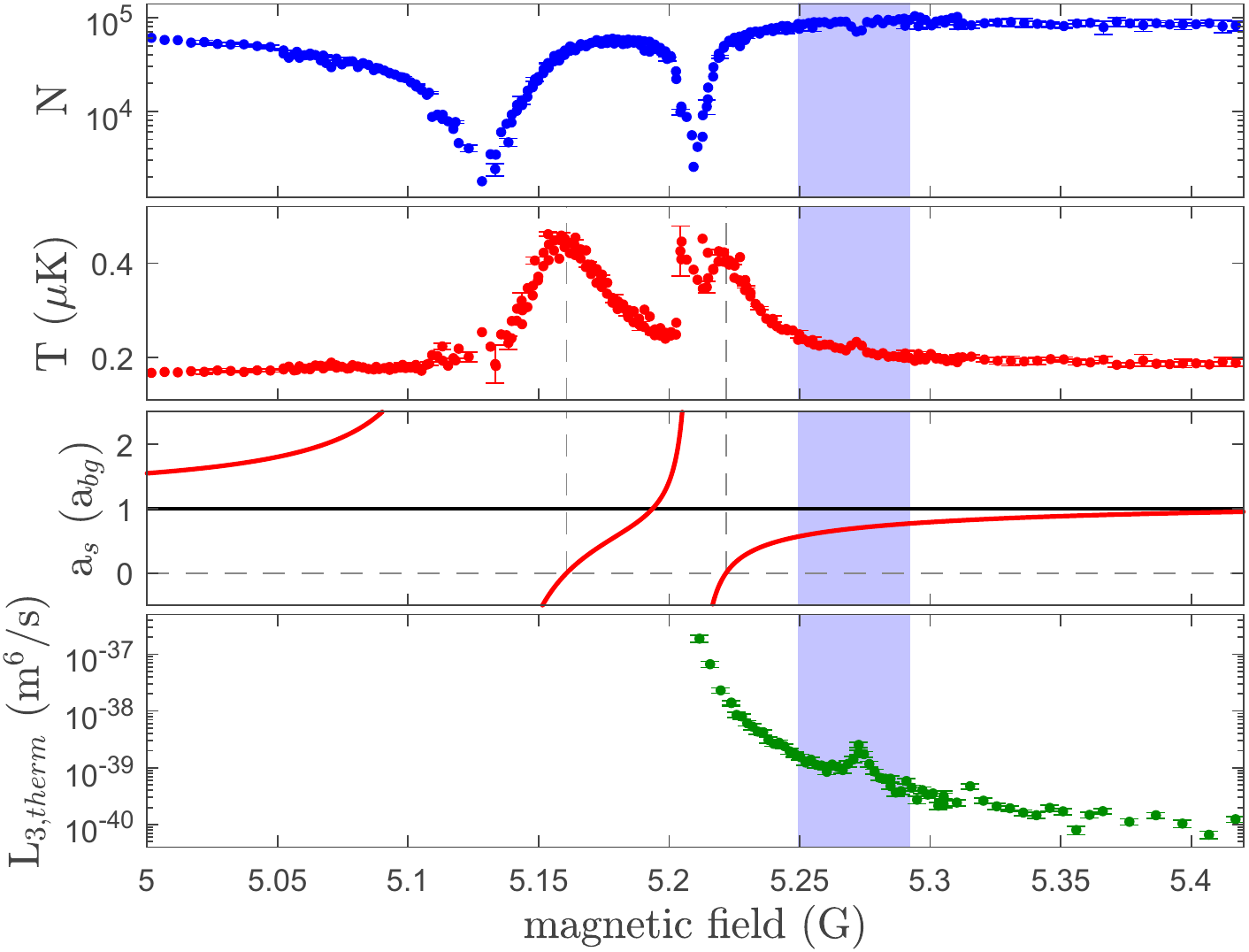}
\put(94, 68){a)} \put(94, 55){b)} \put(94, 38.5){c)} \put(94, 22){d)}
\end{overpic}
\caption{\label{SupMatfig:1}
Combination of Feshbach resonances used to tune the scattering length. We measure the atom number (a) as well as the temperature (b) of a thermal cloud after forced evaporation at different magnetic fields. The extracted positions are $B_{\text{1}}~=~5.126(1)\,\text{G}$ and $B_{\text{2}}~=~5.209(1)\,\text{G}$ with widths of $\Delta B_{\text{1}}~=~35(1)\,\text{mG}$ and $\Delta B_{\text{2}}~=~12(1)\,\text{mG}$, respectively. Together with another resonance at $B_{\text{3}}~=~21.95(5)\,\text{G}$ with a width of $\Delta B_{\text{3}}~=~2.4(8)\,\text{G}$, we can calculate the dependence of the scattering length on the magnetic field (c). The dashed, vertical gray lines represent the position of the zero-crossing of the scattering length, while the blue area corresponds to the region where we observe self-bound quantum droplets. Measured three-body loss coefficient $L_3$ in a thermal cloud (d), which increases the closer we get to the resonance explaining the shorter lifetime of the observed self-bound droplets.
}
\end{figure}

The complicated spectrum of Feshbach resonances for lanthanide atoms allows to control the short-range contact interaction by tuning the amplitude of the magnetic field clos to one of the many resonances. In this work we use a specific double resonance (see Fig.~\ref{SupMatfig:1}) at a field of $B_{\text{1}}~=~5.126(1)\,\text{G}$ and $B_{\text{2}}~=~5.209(1)\,\text{G}$ with widths of $\Delta B_{\text{1}}~=~35(1)\,\text{mG}$ and $\Delta B_{\text{2}}~=~12(1)\,\text{mG}$, respectively. In order to calibrate the magnetic field amplitude we use radio-frequency spectroscopy to get the absolute value of the magnetic field amplitude from the Zeeman shift with an uncertainty of $\sim$1\,mG.  To measure the position and the width of the Feshbach resonances we prepare a thermal cloud with $N~\approx~3.5~\times~10^5$ and $T~\approx~850\,$nK using forced evaporation at a fixed evaporation field ($B~=~5.875\,\text{G}$) and subsequently change the magnetic field within 1\,ms to its final value and further ramp down the intensity of our crossed optical dipole trap in $2\,$s. After 100$\,$ms hold time at the final values we record the atom number (Fig.~\ref{SupMatfig:1}(a)) as well as the temperature (Fig.~\ref{SupMatfig:1}(b)) of the cloud after time of flight. From this we can extract the position of the Feshbach resonances as well as the position of the zero-crossings of the scattering length. The two mentioned resonances, as well as a third resonance, can be seen in Fig.~\ref{SupMatfig:1}. The third resonance at 5.273\,G has a width of only $\sim$1\,mG. This resonance seems to vanish at even lower temperatures and therefore does not influence the atoms in a BEC. Because of this we do not include this narrow resonance in our consideration of the scattering length. Additionally there is a broader resonance at $B_{\text{3}}~=~21.95(5)\,\text{G}$ with a width of $\Delta B_{\text{3}}~=~2.4(8)\,\text{G}$ \cite{Lucioni2018}, that still has a small effect on the scattering length in the magnetic field range considered in this work. Using the mentioned resonances we can calculate the scattering length as a function of the magnetic field, with only the background scattering length as a free parameter (Fig.~\ref{SupMatfig:1}(c)).

On top of the atom-loss spectroscopy used to extract the position and the width of the Feshbach resonances, we also need to measure the three-body loss rate $L_3$ and check whether the observed shorter lifetimes of the $^{162}$Dy droplets can be understood from the theory. To measure $L_3$ we prepare a thermal cloud at about 200\,nK and then ramp up the magnetic gradient, and again compensate the magnetic field shift by ramping up the offset field at the same time. After this we compress the atomic cloud by ramping up the powers in the cODT within 25\,ms, such that we reach a trap with trap frequencies of (83(4), 299(3), 434(2))\,Hz. Then we change the magnetic field to its variable final value within 3\,ms and subsequently let the atoms evolve for up to 1\,s. We then image the atoms after 10\,ms of time of flight and fit the atom number $N$ and the temperature $T$ in order to extract the three-body loss rate to the differential equations, similarly to the methods described in \cite{Weber2003}.
\begin{equation}
\frac{dN}{dt} = - \alpha N - \gamma \frac{N^3}{T^3} \quad \text{and} \quad \frac{dT}{dt} = \gamma \frac{N^2 (T+T_h)}{3\,T^3}.
\end{equation}
In these equations $\alpha$ is the two-body loss rate, which we measured in a dilute thermal cloud to be $\sim$20\,s and $\gamma$ is connected to the three-body coefficient $L_3$ via
\begin{equation}
\gamma = \frac{L_3}{ \sqrt{27}} \, \left(\frac{ m \bar{\omega}^2 }{2 \pi k_B} \right)^3.
\end{equation}
Here $\bar{\omega}$ is the mean trap frequency, $m$ is the mass of the atoms and $k_B$ is the Boltzmann constant. The temperature of the sample increases due to the losses, because two of the colliding atoms can form a molecule and the third can gain the binding energy.

What we see in the last row of Fig.~\ref{SupMatfig:1} is a three-body loss coefficient $L_3~=~8~\times~10^{-41}\,\text{m}^6/\text{s}$ away from the Feshbach resonances, that increases the closer we get to the resonance. Compared to the thermal cloud, the three-body loss coefficient in the BEC is decreased by a factor of 6 \cite{Burt1997}, leading to $L_3~=~1.33~\times~10^{-41}\,\text{m}^6/\text{s}$ away from the resonances. In the droplet region (indicated by the blue area in Fig.~\ref{SupMatfig:1}) we have an $L_3$ that is a factor of 4 ($a_{\text{s}}~\sim~105\,a_0$) or even 15 ($a_{\text{s}}~\sim~80\,a_0$) times larger than the value far away from the resonance. In this range we also see a small peak due to the very narrow Feshbach resonance located there, which we however do not observe any more for lower temperatures. The observed large increase of $L_3$ explains the shorter lifetime that we observe for the self-bound droplets for lower scattering lengths.

For the comparison to the $^{164}$Dy data from \cite{Schmitt2016a}, we convert the given magnetic field amplitudes using the Feshbach resonances at $B_{164,1}$~=~7.117(3)\,G with a width of $\Delta B_{164,1}$~=~51(15)\,mG and a second resonance at $B_{164,2}$~=~5.1(1)\,G with a width of $\Delta B_{164,2}$~=~0.1(1)\,G.

\section{Extended Gross–Pitaevskii simulations} 

We compare our experimental results to theory \cite{Baillie2016, Wachtler2016a}, by performing numerical simulations of the extended Gross-Pitaevskii equation (eGPE) 

\begin{eqnarray}
i \hbar \partial_t \Psi(\vec{r},t) &=& \left[ - \frac{\hbar^2 \nabla^2}{2 m} + V_{\text{ext}} + g \, |\Psi|^2 - i \, \frac{\hbar L_3}{2} \, |\Psi|^4 \right. \nonumber \\*
 \label{eq:eGPE} &\quad +&  \int V_{\text{dd}}(\vec{r}-\vec{r'}) \, |\Psi(\vec{r'})|^2 \, d\vec{r'} \\*
&\quad +& \left. \frac{32 \, g \, \sqrt{a_{\text{s}}^3}}{3 \sqrt{\pi}} \; \left( 1 + \frac{3}{2} \, \varepsilon_{\text{dd}}^2 \right) \, |\Psi|^3 \right] \; \Psi(\vec{r},t)  \nonumber 
\end{eqnarray}
using a simple interaction potential and taking the quantum fluctuations and three-body losses into account within a local density approximation. In this equation $g = 4\pi \hbar^2 \, a_{\text{s}} / m$ is the contact interaction parameter, 
\begin{eqnarray}
V_{\text{dd}}(\vec{r}) = \frac{\mu_0 \mu^2}{4 \pi} \; \frac{1 - 3 \, \cos^2(\theta) }{|\vec{r}|^3}
\end{eqnarray}
is the dipole-dipole interaction of polarized particles, with $\theta$ being the angle between the polarization direction and the relative orientation of the dipoles, and $\mu = 9.93\,\mu_{\text{B}}$ is the magnetic dipole moment of $^{162}$Dy. We change the scattering length $a_{\text{s}}$ in the range of 60\,$a_0$ to 115\,$a_0$. Furthermore, we use the measured $L_3$ parameter for the respective scattering lengths. The eGPE in Eq. \ref{eq:eGPE} uses two assumptions: the Born approximation for the interaction potential and the local density approximation. The second approximation is supported by quantum Monte-Carlo simulations \cite{Saito2016} and a comparision of theory and experiment with erbium \cite{Chomaz2016a}. The first assumption is studied in \cite{Odziejewski2016, *Odziejewski2017Erratum} and needs to be adjusted at finite temperature. The departure from the Born approximation can be taken into account by an effective dipolar length $a_{\text{dd}}$ that is shifted by a few percent compared to the dipolar length within the Born approximation.

In order to get the theory curve in Fig.~\ref{fig:critNum} we used two different methods leading to the same result. For both methods we choose $V_{\text{ext}}~=~0$ and start with an atom number $N~>~N_{\text{crit}}$, initially prepared with an elongated Gaussian density distribution. Then we find the ground state by imaginary time evolution of the eGPE. Next we can either simulate atom losses like in the experiment, or we can repeat this process of finding the ground state with ever lower atom number to start with until we do not find a stable solution anymore. In the second method we get an uncertainty due to the step size that we choose for the atom number. For the first method we do real-time evolution of the eGPE in order to simulate the dynamics of three-body losses. Due to the losses, the density and the effective two-body attraction reduces with time, until we reach $N~=~N_{\text{crit}}$ where the contributions of the effective two-body attraction and the quantum pressure cancel each other. This leads to the evaporation of the droplet into the gaseous phase, which leads to a steep decrease of the density by more than an order of magnitude. This suppresses further losses and the atom number stays almost constant as soon as the droplet has evaporated, validating the interpretation of the experimentally observed loss curves.

\section{Droplet phase diagram}

\begin{figure}[t]
\begin{overpic}[width=0.32\textwidth]{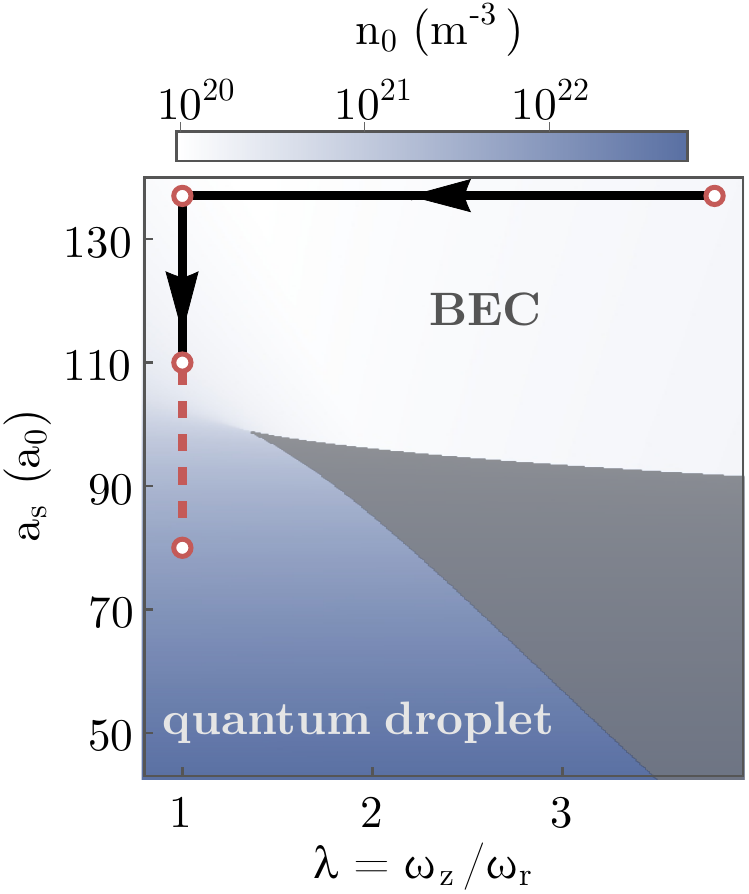}
\end{overpic}
\caption{\label{SupMatfig:2}
Phase diagram and schematic steps of the experiment. The phase diagram is calculated using the Gaussian ansatz to solve the eGPE for a cylindrically trapped $^{162}$Dy BEC, with mean trap frequency $\bar{\omega}~=~\sqrt{\omega_{\text{x}}\omega_{\text{y}}\omega_{\text{z}}}~=~30\,$Hz and containing $2~\times~10^{4}$ atoms. For the trap aspect ratio $\lambda$ below the critical point  $\lambda~<~\lambda_{\text{c}}$, the stable BEC solution and the single quantum droplet state are connected through a continuous crossover. Above the critical point $\lambda_{\text{c}}$ there is a multi-stable region where both solutions are stable (shown in gray).
}
\end{figure}

We can calculate the phase diagram \cite{Bisset2016, Wachtler2016a} of trapped dipolar atoms as a funtion of the trap aspect ratio $\lambda$ and the scattering length $a_{\text{s}}$ using the eGPE, and then either apply a Gaussian ansatz to solve it analytically or resort to full numerical simulations. Here, we are only interested in a qualitative discussion and therefore restrict ourself to the Gaussian ansatz, the full simulations together with a measurement of the critical point $\lambda_{\text{c}}$ can be found in \cite{Ferrier-Barbut2018}. The calculated phase diagram for a cylindrically trapped BEC, with mean trap frequency $\bar{\omega}~=~\sqrt{\omega_{\text{x}}\omega_{\text{y}}\omega_{\text{z}}} = 30\,$Hz and containing $20,000$ $^{162}$Dy atoms is shown in Fig.~\ref{SupMatfig:2}. 

The phase diagram contains three different regions: For large scattering lengths only a single repulsive BEC solution of the eGPE exists (white region in Fig \ref{SupMatfig:2}), which has a cloud aspect ratio close to that of the trap only weakly altered by magnetostriction \cite{Stuhler2007}. At low scattering lengths, also only a single solution exists, with a cloud aspect ratio $\ll 1$ that is more or less independent of $\lambda$. This is the quantum droplet solution (blue region) that is only stable due to beyond mean-field corrections \cite{Ferrier-Barbut2016, Chomaz2016a, Bisset2016, Wachtler2016a}. This solution has a peak density that is a factor of $\gtrsim 10$ higher than that of the BEC. For trap aspect ratios larger than $\lambda_{\text{c}}$ there exists a bistable region (gray region) where both solutions can be stable. Crossing the boundary leads to a modulational instability and therefore higher number of droplets than what is expected for the ground state of the system. In the case of $\lambda~<~\lambda_{\text{c}}$ the two phases are connected through a continuous crossover instead of the phase transition in the bistable region. 

In Fig.~\ref{SupMatfig:2} we also show the way in which we prepare the single droplet ground state in the experiment (indicated by the red circles and the black arrows). We start with the atoms in our cODT with a trap aspect ration $\lambda~=~3.8$ and a scattering length $a_{\text{s}}~\approx 140\,a_0$, deep in the BEC regime. We then change the trap aspect ratio to $\lambda~\approx~1$ and then change the magnetic field amplitude to probe the crossover to the droplet phase. In the droplet regime we can then turn off the trap completely and observe the self-bound state in free space.

\section{Expansion velocity}

\begin{figure}[t]
\begin{overpic}[width=0.47\textwidth]{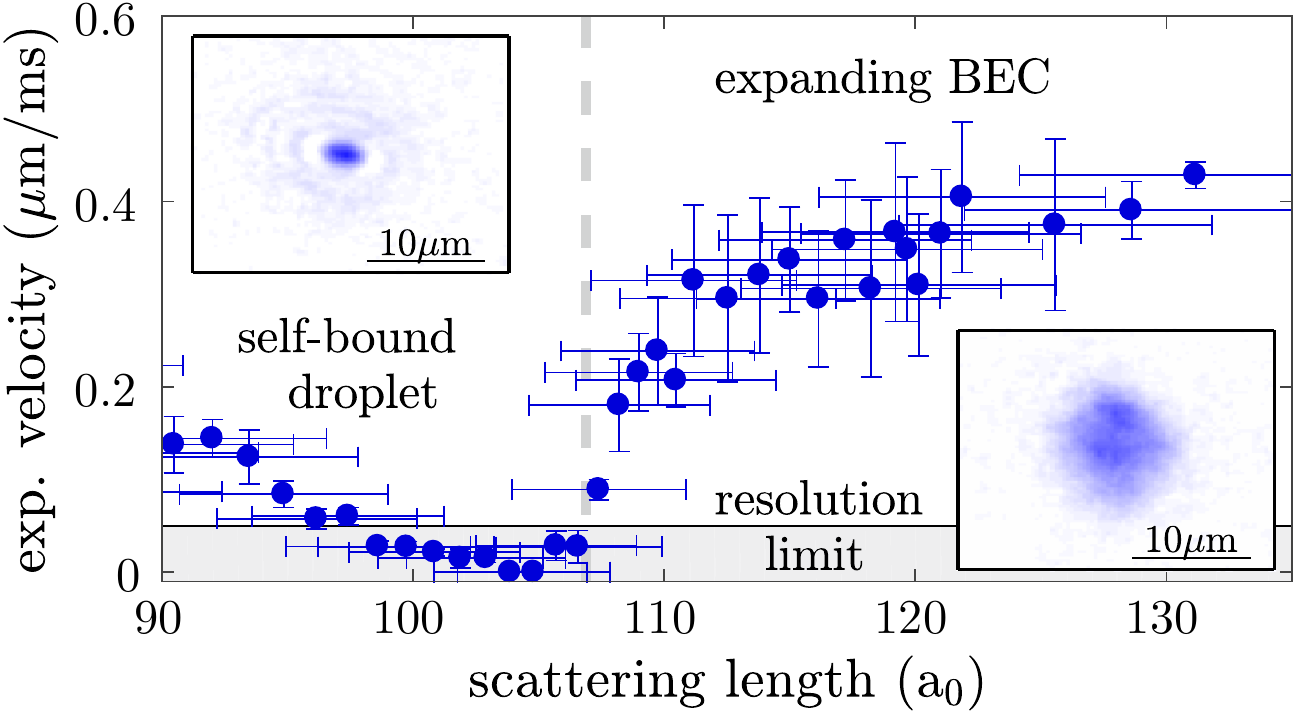}
\end{overpic}
\caption{\label{SupMatFig:ExpVelo} 
Expansion velocity across the crossover from a BEC to a quantum droplet. We extract the expansion velocity $v_{\text{exp}}$ from the evolution of the widths of the atomic cloud in time-of-flight for up to 20\,ms, averaged over 5 realizations. This procedure can be applied both in x and y direction, for which we find comparable results. We thus plot here the average over both directions. As errorbars we show the quadratic sum of the uncertainty of the determination of $v_{\text{exp}}$ along the two directions and for the scattering length the uncertainty due to the experimental field stability, the knowledge of the Feshbach resonances, as well as the background scattering length.  The two insets show example single-shot images for a self-bound droplet (top) and an expanding BEC (bottom) after 20$\,$ms time of flight.
}
\end{figure}

While a BEC is a gaseous state, which means that it freely expands in the absence of an external trap, the droplet state is self-bound due to its intrinsic interactions \cite{Baillie2016,Wachtler2016a, Schmitt2016a} and therefore does not expand. To map out the range of self-bound droplets, we determine the expansion velocity $v_{\text{exp}}$ \cite{Chomaz2016a} by fitting the evolution of the observed widths of the atomic cloud $\sigma_{\text{tof}}$ up to $t_{\text{tof}} = 20\,ms$ to $\sigma_{\text{tof}}~=~\sqrt{\sigma_{\text{0}} ^2 + v_{\text{exp}}^2 \, t_{\text{tof}}^2}$. In this $\sigma_{\text{0}}$ corresponds to the size at zero time-of-flight. The extracted expansion velocity across the complete crossover from stable BEC to single droplet state is shown in Fig.~\ref{SupMatFig:ExpVelo}. Entering the regime with $\varepsilon_{\text{dd}}~>~1$ we observe a small decrease of the expansion velocity with higher relative dipolar strength, until at around $a_{\text{s}} \approx 110\,a_{0}$ where we observe a sharp decrease. For $a_{\text{s}} \lesssim 107 a_{\text{0}}$ we enter the self-bound regime, where we do not observe an increase in the size within our 1$\,\upmu$m imaging resolution. In this regime we also observe aberrations in the images (see top inset in Fig.~\ref{SupMatFig:ExpVelo}) due to the small radial size compared to our imaging resolution. After some time, depending on the magnetic field and on the initial atom number, we find that the cloud has expanded even in the droplet state. This is understood in terms of three-body decay until it reaches the critical number below which the droplet is not self-bound anymore. The short lifetime due to enhanced three-body losses closer to the resonances leads to the observed increase of the expansion velocity for lower scattering lengths. 

\section{Experimental determination of the critical atom number}

\begin{figure}[t]
\begin{overpic}[width=0.48\textwidth]{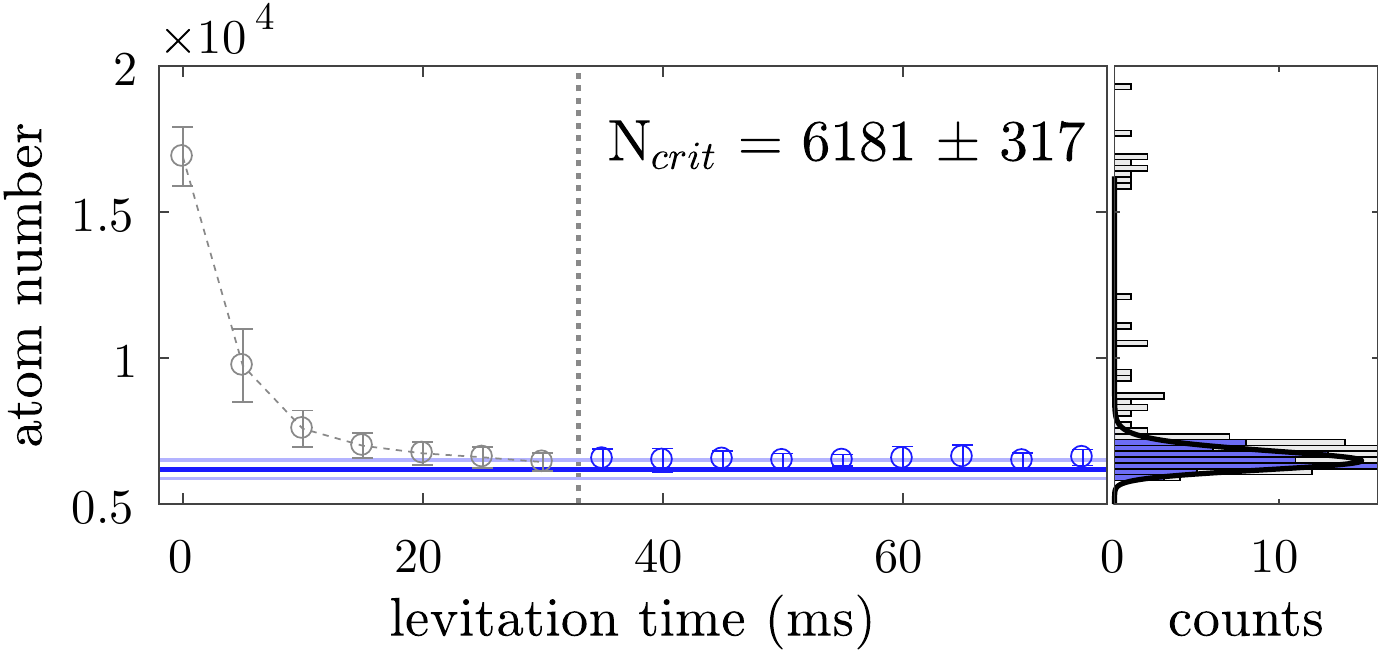}
\end{overpic}
\caption{\label{SupMatfig:4}
Exemplary atom number decay curve of a self-bound droplet. Measured decay of the atom number as a function of the levitation time, averaged over 10 realizations and the error bar denote the respective standard deviations. After a fast decay for short times (gray points), we observe a constant atom number (blue points). To extract the critical atom number $N_{\text{crit}}$ (horizontal blue line) we analyze the atom number distribution (histogram on the right side) and fit our convolution model to the blue data (black line).
}
\end{figure}  

In order to measure the critical atom number of a self-bound droplet we look at atom number decay curves \cite{Schmitt2016a}, as exemplary shown in Fig.~\ref{SupMatfig:4} for a scattering length of $a_{\text{s}} = 99(7)\,a_{0}$. We use a sequence of intentional evaporation and subsequent expansion that allows us to determine the atom number precisely without being limited by the finite resolution of our imaging system or by the high density of the droplets. This intentional evaporation is done by ramping up the magnetic field within 100\,$\upmu$s to $B_{\text{evap}} \approx 6.0\,$G. We know that the ground state of the system is a BEC for this magnetic field. Therefore, by quickly ramping up the field, we force the droplet back into a gaseous state, where it then expands freely. We let the atomic cloud expand for either 8\,ms or 30\,ms, depending on the atom number, and then we image the atomic cloud with resonant absorption imaging. We use this sequence for different magnetic fields in the range from $B = 5.293\,\text{G}$ to $B = 5.249\,\text{G}$, corresponding to $a_{\text{s}} = 107\,a_0$ and $a_{\text{s}} = 78\,a_0$, respectively. 

With this sequence we observe that the atom number decays very fast in the beginning, but then settles to a constant value. This behaviour results from an initial fast three-body decay of the high-density droplet state, followed by a quick expansion of the droplet as it crosses the phase boundary to the gaseous state. The crossing of the phase boundary leads to a fast drop in density and thus suppresses further loss. To extract the critical number we employ a statistical evaluation procedure, because the critical atom number is reached at different times during these 10 realizations due to the stochastic preparation of the initial conditions in our experiment. We show histograms of the atom number distribution on the right side of Fig.~\ref{SupMatfig:4}. The first histogram includes all data points (gray bars), while the second histogram only includes the points for which the atom number has settled to its final value (blue bars). 

We use a phenomenological model (black line in the histogram of Fig.~\ref{SupMatfig:4}) in order to extract the critical atom number of a self-bound droplet \cite{Schmitt2016a}. This model consists of a convolution of a Gaussian and a Maxwell–Boltzmann distribution. We use a symmetric Gaussian distribution in order to represent broadening effects that result from statistical errors, e.g. noise in the imaging of the atomic cloud. To cover the possibility that residual collective excitations in the droplet lead to an early evaporation at atom numbers higher than the critical number, we use an asymmetric Maxwell-Boltzmann distribution. By fitting this phenomenological model we extract the critical atom number and two different widths, one for each distribution. As an error bar of the critical atom numbers in Fig.~\ref{fig:critNum} of the main text we use the quadratic mean of these two widths, together with an overall 10\% uncertainty due to the uncertainty in calibration of the imaging system.

\section{Experimental determination of the background scattering length}

\begin{figure}[t]
\begin{overpic}[width=0.48\textwidth]{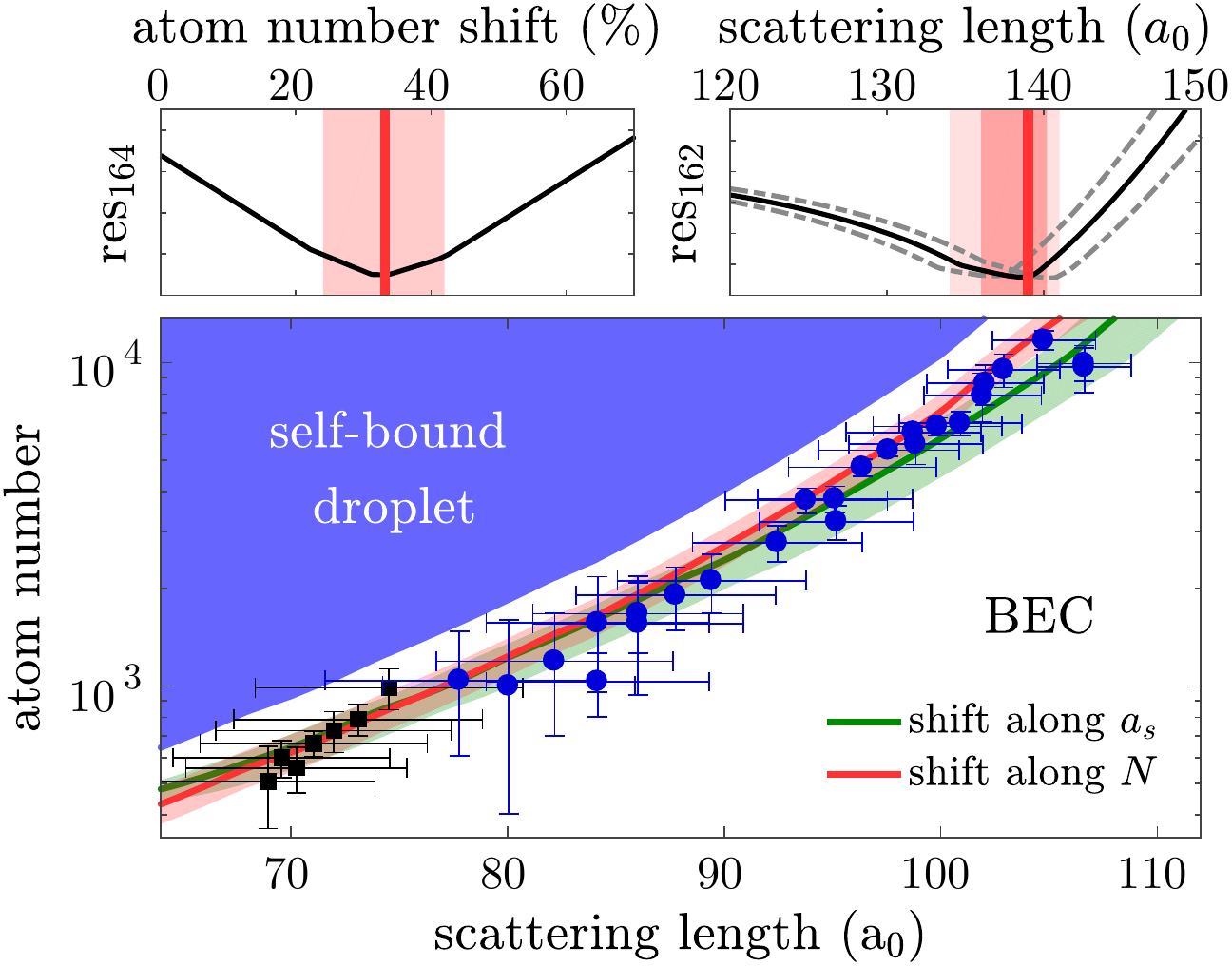}
\put(14.5,66){a)} \put(61,66){b)} \put(14.5,48.5){c)}
\end{overpic}
\caption{\label{SupMatfig:5}
Determination of the background scattering length of $^{162}$Dy. a) Difference res$_{\text{164}}$ between measured critical atom numbers for $^{164}$Dy and the results obtained from numerical simulations shifted along the atom number axis. The vertical red line indicates the shift with the lowest residual from the shifted theory and the lighter red area marks the range of uncertainty upon which the residual doubles. b) Residual difference between the observed critical atom numbers of $^{162}$Dy and the shifted theory curve (black) with the minimum value shown by the vertical red line. The two dashed lines represent the residual using the boundaries of the uncertainty of the shifted theory. c) Summary of measured Critical atom number versus scattering length. The red line shows the shifted theory curve for a shift along the atom number axis and the green line similarly for a shift along the scattering length axis. The lighter areas show the respective uncertainty of the shift. 
}
\end{figure}  

Using the latest measurement of the background scattering length of $^{164}$Dy \cite{Ferrier-Barbut2018a} as a starting point, we use the sensitive scaling of the critical atom number to extract the background scattering length of $^{162}$Dy. This procedure, that is described in detail below, assumes that the critical atom number does not depend on other parameters, e.g. the actual density distribution of the droplet. 

Since we observe a systematic shift of our measured critical numbers compared to the results obtained from numerical simulations, we first shift the theoretical curve in order to minimize the difference (res$_{\text{164}}$) between the experimental data for $^{164}$Dy \cite{Schmitt2016a} and the shifted theory. This shift of the simulated boundary between self-bound droplet and expanding BEC can be done along the atom number axis or along the scattering length axis, resulting in slightly different results. In Fig.~\ref{SupMatfig:5}(a) we show the obtained residual for a shift along the atom number axis. The curve presents a clear minimum for a shift of 30\% of the corresponding simulated critical atom numbers to lower values (marked by the vertical red line). As uncertainty we use the range where this residual from the shifted theory has doubled (shown as light red).

From this point on, we then optimize the background scattering length $a_{\text{bg,162}}$ such that we get the smallest residual of the measured critical atom numbers for $^{162}$Dy with respect to the shifted theory curve. This can be done since with our knowledge of the parameters of the used Feshbach resonances the background scattering length is the only free parameter not known precisely. In Fig.~\ref{SupMatfig:5}(b) we show the residual res$_{\text{162}}$ for the case of a shifted theory along the atom number axis. From this we get two uncertainties, one again extracted from a doubling of the corresponding residual (light red area) and the second from the propagation of the uncertainty from the shifted theory curve (darker red area). With this procedure with the shift along the atom number axis we get a value of $a_{\text{bg,162}} = 139(135, 141)\,a_0$ for the background scattering length of $^{162}$Dy. 

Similarly this can be done for a shift along the scattering length axis resulting in $a_{\text{bg,162}} = 141(136, 145)\,a_0$. Taking the average of these two procedures we end up with $a_{\text{bg,162}} = 140(4)\,a_0$, in good agreement with the quoted literature value \cite{Tang2018, Tang2015a, Tang2016}. Note, however, that the systematic uncertainties for the measurement of the background scattering length of $^{164}$Dy \cite{Ferrier-Barbut2018a} apply here as well, leading to a final value with the respective uncertainty of $a_{\text{bg,162}} = 140(7)\,a_0$.

Free from the systematic shift of the measurement, we can extract the ratio of the respective background scattering lengths of the two isotopes, $^{162}$Dy and $^{164}$Dy. With this we find a ratio of $a_{\text{bg,162}}/a_{\text{bg,164}} = 2.03(6)$.

\section{Effective re-normalization of the dipolar length}

One possible explanation of lower critical atom numbers of a self-bound droplet arises due to the complexity of the scattering problem for dipolar lanthanide atoms such as dysprosium.  As it was pointed out in \cite{Odziejewski2016, *Odziejewski2017Erratum} taking into account the full scattering amplitude leads to alterations compared to the Born approximation. This can be accounted for within the eGPE framework by an effective shift in the dipolar length $a_{\text{dd}}$ that depends on the collision energy. This effective shift is on the order of 2\% for a temperature of 10\,nK and 10\% for 100\,nK. As our BEC has an initial temperature of about 15\,nK and the preparation process likely leads to additional heating, it may seem reasonable to include this effect in our considerations. Since the droplets are self-bound, we cannot use standard time of flight expansion to measure the temperature of these states. In the future, we plan to resort to novel methods to measure the temperature inside the droplets, e.g. with embedded impurities \cite{Wenzel2018} similar to liquid helium droplets \cite{Toennies2004}. With this, we can also clarify the role played by thermal fluctuations at larger atom numbers \cite{Aybar2018, *Boudjemaa2017}.

In Fig.~\ref{fig:critNum} we also show two theory curves for an enhanced dipolar length: $a_{\text{dd}} + 5\%$ enhancement (dashed red line) corresponding to a temperature of 30-50\,nK \cite{JachymskiPrivateComm}, and $a_{\text{dd}} + 10\%$ enhancement (dash-dotted red line) corresponding to 100\,nK. The observed theoretical shifts are in good agreement with the experimental shifts, suggesting that such an effective enhancement of the dipolar length might play an important role in dipolar scattering at finite temperature.

\section{Path Integral Ground State}
Given a Hamiltonian, the Path Integral Ground State (PIGS) method can be used to evaluate \textit{exactly} many-body properties of a correlated Bose system, beyond the mean field plus Lee-Huang-Yang approximation. Designed as a reduction of the Feynman path integral formalism to zero temperature, particle coordinates at different (but close) imaginary times are sampled in chains, starting from a variational wave function $\Psi_T(R)$ that is located at the end points. Since propagation in imaginary time removes any component that is orthogonal to the true ground-state wave function $\phi_0(R)$ in the asymptotic limit, samples of $\phi_0(R)$ are realised at the center of the chains when the total propagation time is long enough

\begin{eqnarray}
\label{eq.PIGS}
\phi_0(R_M) & = & \lim_{\substack{\delta \tau\to 0 \\ M\to \infty}}\int dR_ {M-1}... dR_1 \\
& & \times \prod_{i= 1}^{M-1}G(R_{i+1},R_{i},\delta\tau)\Psi_T(R_1) \ . \nonumber 
\end{eqnarray}
In this expression, $G(R_{i+1},R_{i})$ is the imaginary time propagator between positions R$_{i+1}$ and R$_i$ in a time step $\delta\tau$, which is related to the action $\hat{S}$ through the expression $\hat{G} = e^{-\hat{S}}$.  In general, the action $\hat S$ is not known, but since $\delta\tau$ is small, a low-order series expansion in powers of $\delta\tau$ can be successfully employed. In this work we use one of the fourth-order propagators of Ref.~\cite{Chin2002}, which improves convergence when compared with other, simpler schemes based on a second-order Trotter expansion.

The main ingredient required to perform a PIGS simulation is the Hamiltonian. Here, we use a model that includes both the dipolar interaction and an effective potential $V_{HC}$ with a repulsive core that prevents the system from collapsing. Assuming that all the dipoles are polarized along the Z axis, the Hamiltonian reads
\begin{equation}
\hat{H}=-\frac{\hbar^2}{2m}\sum_{i=1}^N\nabla^2_i+\frac{C_{dd}}{4
  \pi}\sum_{i<j}\frac{1-3cos^2\theta_{i,j}}{r_{i,j}^3} + V_{HC} +
V_{trap},
\label{eqn:hamiltonian}
\end{equation}
where $\vec{r}_{i,j}$ and $\theta_{i,j}$ are the relative polar coordinates between the atoms, $m$ is the atomic mass, and $C_{dd}$ sets the strength of the dipolar interaction. In order to study whether there are universality properties in the system at the given conditions, we solve the Hamiltonian for three different $V_{HC}$ models

\begin{figure}[h]
\begin{center}
\includegraphics[width=9cm]{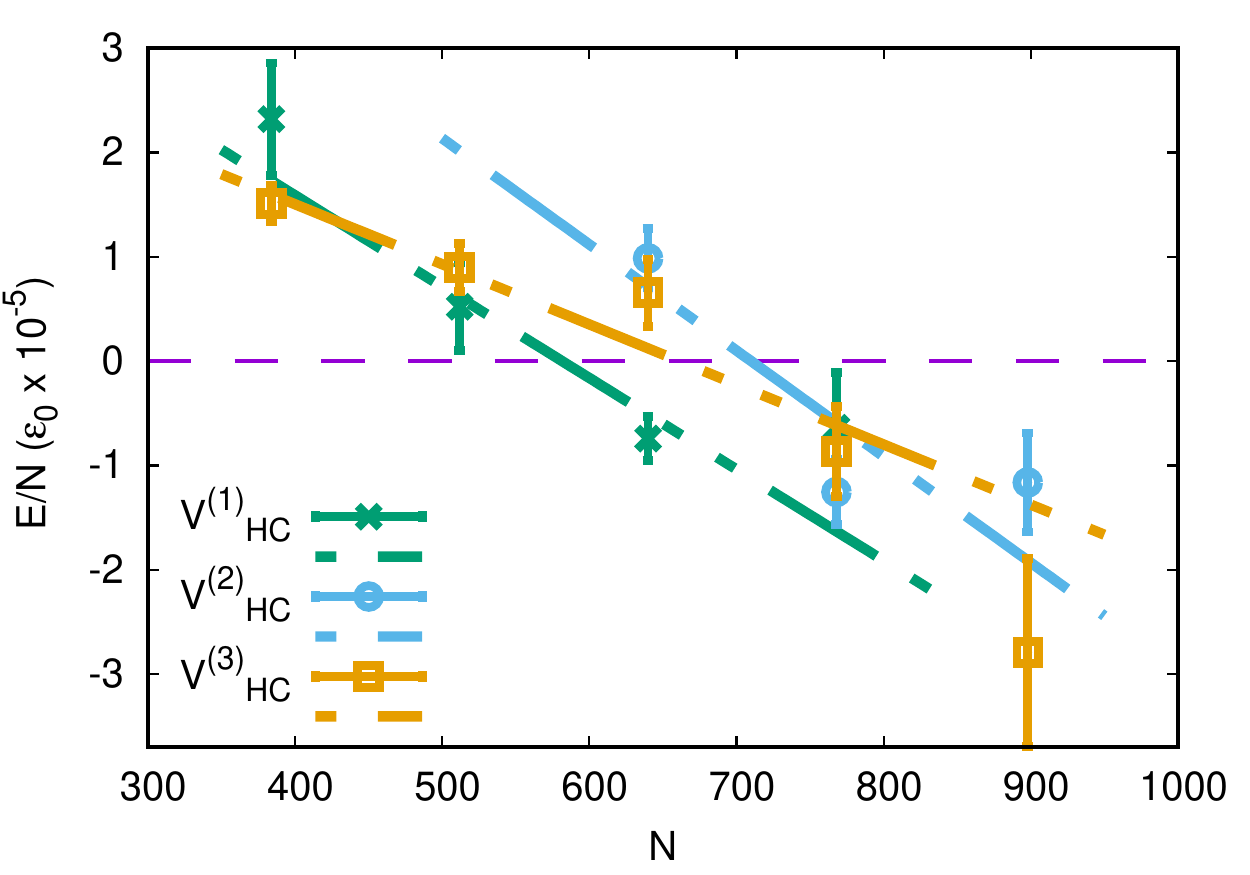}
\caption{Energy per particle in units of $\hbar^2/m a_{dd}^2$ for the dipolar system with the three interaction of Eq.~(\ref{eqn:potentials}) for the $s$-scatering length $a_s$ = 60$a_0$. The lines represent a fit to the data, and the intersection with the $E=0$ axis defines the critical number of the model at this scattering length value.}
\label{fig:non_univ}
\end{center}
\end{figure}

\begin{eqnarray}
  V_{HC}^{(1)} & = & \left( {\sigma_{12} \over r} \right)^{12} -
  { C_6 \over r^6 } \nonumber \\
  V_{HC}^{(2)} & = & \left( {\sigma_{9} \over r} \right)^{9} -
  { C_6 \over r^6 } \nonumber \\
  V_{HC}^{(3)} & = & \left( {\sigma_{12} \over r} \right)^{12}
\label{eqn:potentials}  
\end{eqnarray}
The coefficient $C_6$ in the previous equation is known for Dysprosium~\cite{Li2017}. The other coefficients, $\sigma_9$ and $\sigma_{12}$, are fixed such that the complete interaction ($V_{HC}$ plus dipolar) has the experimental $s$-wave scattering lengths. This is accomplished by solving the low momentum limit of the scattering T-matrix, as briefly described in the next section.

One of the fundamental quantities that can be obtained from the PIGS simulations is the ground state energy, which is negative for a self-bound droplet state. As in the experiments and for a given scattering length, we find that there is a critical number $N_c$ below which the system ceases to be self-bound. Fig.~\ref{fig:non_univ} shows, for $a_s=60 a_0$, the ground-state energy obtained from the Hamiltonian in Eq.~(\ref{eqn:hamiltonian}) for the three $V_{HC}$ models, as a function of the total number of particles, together with a linear fit that determines the point $N_c$ where the energy is zero. As it can be seen, different models lead to slightly different predictions, and that defines, together with the statistical noise in the simulation, the error bar in the critical number $N_c$. As the scattering length increases, higher values of both $N_c$ and its error are found, but they are all compatible with the experimental error bars. Unfortunately, the computational cost of the simulation grows very fast with the number of particles and we can not reliably determine $N_c$ for scattering lengths larger than $a_s=90 a_0$, approximately.

\begin{figure}[t]
\begin{center}
\includegraphics[width=9cm]{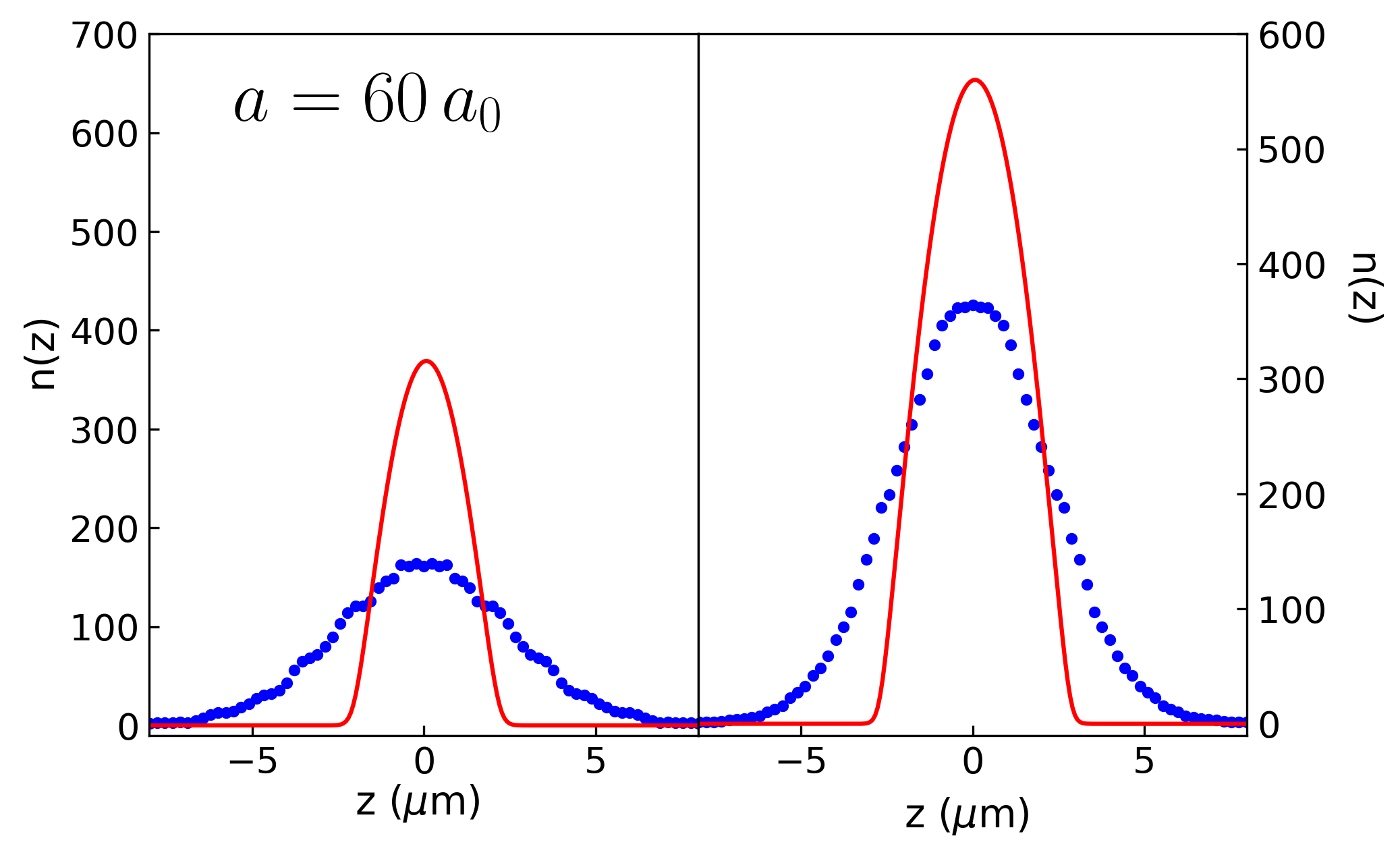}
\caption{Density profiles along the $z$ direction in the eGPE (red solid line) and PIGS (blue dots) approximations for a scattering length $a=60 a_0$. The left and right panels show the eGPS results for $N=1000$ and $N=2000$ atoms, compared with the PIGS results for $N=1024$ and $N=2048$ atoms, respectively. Each profile has been properly normalized to its corresponding particle number. }
\label{fig:dens_prof_degp_pigs}
\end{center}
\end{figure}

The droplets predicted by PIGS differ from those obtained in the eGPE approximation not only in the critical number, but also on the overall density profiles. Figure~\ref{fig:dens_prof_degp_pigs} shows the integrated density profiles along the axial directions of the droplet, obtained from both methods, for $~1000$ (left panel) and $~2000$ (right panel) atoms, and for a scattering length of $60 a_0$. As it can be seen, for these (low) particle numbers, the profiles are different, the PIGS one being more spread but with a lower central density. Still, the difference reduces when the number of atoms is increased from 1000 to 2000. Increasing the atom number even more, we expect the differences in the density profile to decrease.

\section{Calculation of the s-wave scattering length for a two-body potential}
\label{ap.Tmatrix}

The $s$-wave scattering length of the combined two-body plus dipole-dipole interaction is obtained from the on-shell $T$-matrix, in the limit of vanishing momentum transfer. The $T$-matrix can be obtained solving the Lippmann-Schwinger equation projected on a basis of free-particle eigenstates of definite angular momentum, according to
\begin{eqnarray}
 T_{l',m'}^{l,m}(k',k) & = &
 V_{l',m'}^{l,m}(k',k) \label{Lippman-Schwinger} \\ &+&
 \frac{\hbar^2}{M} \sum_{l_2,m_2} \int \frac{
   V_{l',m'}^{l_2,m_2}(k',q) T_{l_2,m_2}^{l,m}(q,k) }{ \left(
   \frac{\hbar^2 k^2}{2M} - \frac{\hbar^2 q^2}{2M} + i\epsilon \right)
 } q dq \ , \nonumber
\end{eqnarray}
with $V_{l',m'}^{l,m}$ the matrix elements of the complete interaction, and $M$ the reduced mass of two atoms.  Due to the anisotropy of the dipolar potential, the matrix elements of $T$, for different values of the quantum number $l$ and $l'$, are coupled. Moreover, the long-range character of the combined potential makes all partial waves contribute significantly, even at low scattering energies~\cite{Lahaye2009}. Due to the nature of the dipolar interaction, different scattering lengths corresponding to different (coupled) channels appear and read 
\begin{equation}
 a_{l',m}^{l,m} \equiv \lim_{k \rightarrow 0} \frac{\pi T_{l',m}^{l,m}(k,k)}{k}
\end{equation}
with $l'=|l \pm 2|$. Still, the dominant one is the $s$-wave scattering length, corresponding to $l=l'=m=0$.  In practice, the low-momentum matrix elements $T_{l',m}^{l,m}(k,k)$ can be efficiently determined using the Johnson algorithm~\cite{Johnson1973}, which solves the Schr\"odinger equation and finds the logarithmic derivative of the wave function. Table~\ref{tab:table_scattering} shows the $\sigma_\alpha$ parameter of the hard core potentials used in this
work. These values have been chosen such that the resulting interactions do not have any two-body bound state.
\begin{table}[h!]
  \centering
  \begin{tabular}{| l | l | l | l |}
  \hline
  $a_{0,0}^{0,0}$ & $V^{(1)}_{HC}$ & $V^{(2)}_{HC}$ & $V^{(3)}_{HC}$ \\
  \hline\hline
  $60 a_0$ & 1.32 & 1.36 & 1.23 \\
    \hline
  $70 a_0$ & 1.36 & 1.39 & 1.27 \\
    \hline
  $80 a_0$ & 1.41 & 1.43 & 1.34 \\
  \hline     
  $90 a_0$ & 1.46 & 1.48 & 1.40 \\
  \hline 
\end{tabular}
\caption{Values for the parameter $\sigma_\alpha$ of the potentials in Eq.~(\ref{eqn:potentials}), in dipolar units, for different scattering lengths.}
\label{tab:table_scattering}
\end{table}

\bibliography{paper} 

\end{document}